\documentclass[prl,twocolumn,showpacs,preprintnumbers,amsmath,amssymb]{revtex4}

\usepackage{graphicx}% Include figure files
\usepackage{dcolumn}% Align table columns on decimal point
\usepackage{bm}% bold math
\usepackage{amssymb}
\usepackage{float,afterpage,wrapfig}

\DeclareMathAlphabet{\bi}{OML}{cmm}{b}{it}
\newcommand{\dc}{\delta_c}
\newcommand{\zp}{\eta}
\newcommand{\be}{\begin{equation}}
\newcommand{\bigqq}{\bi{Q}}
\newcommand{\bra}{\langle}
\newcommand{\chiqw}{\chi(\bi{q},\w)}

\newcommand{\chiQw}{\chi(\bi{Q},\w)}
\newcommand{\chiQz}{\chi(\bi{Q},0)}
\newcommand{\chiQzi}{\chi^{-1}(\bi{Q},0)}

\newcommand{\cl}{\chi_{\mathrm{loc}}}
\newcommand{\czi}{\chi_0^{-1}}
\newcommand{\dcA}{\delta_{c,\mathrm{A}}}

\newcommand{\dcP}{\delta_{c,\mathrm{P}}}
\newcommand{\ee}{\end{equation}}

\newcommand{\hloc}{h_{\mathrm{loc}}}

\newcommand{\ket}{\rangle}
\newcommand{\kk}{\bi{k}}
\newcommand{\maf}{m_{\mathrm{AFM}}}

\newcommand{\qq}{\bi{q}}
\newcommand{\ra}{\rightarrow}
\newcommand{\rhoi}{\rho_{\mathrm{I}}}

\newcommand{\tkn}{T^0_{K}}
\newcommand{\tk}{T_{K}}

\newcommand{\w}{\omega}

\begin{document}
\title{Magnetic Quantum Phase Transition in an Anisotropic Kondo Lattice}
\author{Matthew T. Glossop}
\author{Kevin Ingersent}
\affiliation{Department of Physics, University of Florida,
Gainesville, Florida 32611-8440, USA}
\date{24 July 2006; published 27 November 2007}

\begin{abstract}
The quantum phase transition between paramagnetic and antiferromagnetic phases
of the Kondo lattice model with Ising anisotropy in the intersite exchange is
studied within extended dynamical mean-field theory.
Nonperturbative numerical solutions at zero temperature point to a continuous
transition for both two- and three-dimensional magnetism. In the former case,
the transition is associated with critical local physics, characterized by a
vanishing Kondo scale and by an anomalous exponent in the dynamics close in
value to that measured in heavy-fermion CeCu$_{5.9}$Au$_{0.1}$.
\end{abstract}

% PACS:
%   71.10.Hf   Non-Fermi-liquid ground states, electron phase diagrams and
%              phase transitions in model systems
%   71.27.+a   Strongly correlated electron systems; heavy fermions
%   75.20.Hr   Local moment in compounds and alloys; Kondo effect, valence
%              fluctuations, heavy fermions (see also 72.15.Qm Scattering
%              mechanisms and Kondo effect)
%   05.10.Cc   Renormalization group methods

\maketitle

Heavy-fermion materials close to a zero-temperature antiferromagnetic (AFM)
instability manifest a rich variety of non-Fermi-liquid behaviors
\cite{Stewart:01}. By tuning a control parameter (doping, pressure, or magnetic
field), the N\'{e}el temperature can be suppressed to $T=0$. The resulting
quantum phase transition (QPT) separates an AFM metal from a paramagnetic (PM)
metal in which local moments are Kondo-screened by conduction electrons. In the
standard spin-density wave (SDW) theory \cite{hm} for such QPTs, the
important critical modes are long-wavelength order-parameter fluctuations;
non-Fermi-liquid physics arises from scattering of quasiparticles by SDWs.
However, data for a number of heavy fermions are at odds with this picture.
In particular, neutron-scattering for CeCu$_{5.9}$Au$_{0.1}$
\cite{neutrons,Schroder:00} points to a strongly interacting quantum critical
point involving novel local physics \cite{Coleman:01}.

Many of the deviations from the SDW theory can be explained if criticality
in the magnetic ordering renders Kondo physics simultaneously critical,
producing a ``locally critical QPT'' \cite{lcqpt,Zhu:03,Si:05}.
Here, work has focused on the Kondo lattice model, mapped within extended
dynamical mean-field theory (EDMFT) \cite{edmft} onto a self-consistently
determined Bose-Fermi Kondo (BFK) model for a magnetic impurity coupled both
to a fermionic band and to a dissipative bosonic bath representing nonlocal
fluctuations arising from Ruderman-Kittel-Kasuya-Yoshida (RKKY) interactions
between local moments.

The central question we set out to answer is whether the EDMFT captures a
\textit{continuous} QPT. Such a transition is expected to be of SDW type for
three-dimensional (3D) magnetic spin fluctuations, with local moments that are
Kondo-screened below a finite low-temperature scale $\tk^*$ and hence play no
role at the QPT. By contrast, if the fluctuations are two-dimensional
(2D)---for which there is some evidence in CeCu$_{5.9}$Au$_{0.1}$
\cite{Rosch:97}---the scale $\tk^*$ is claimed \cite{lcqpt} to vanish
precisely at the QPT with coexisting long-wavelength and local critical modes.
To confirm the scenarios described above, it must be demonstrated that the PM
and AFM phases terminate at a common value of the control parameter. Numerical
studies employing quantum Monte Carlo impurity solvers have addressed
the particular case of Ising anisotropy believed to be relevant to
CeCu$_{5.9}$Au$_{0.1}$ \cite{Zhu:03,Sun:03}. The character of the QPT remains
contentious \cite{Zhu:03,Si:05,Sun:03}, in part due to the limitations of
quantum Monte Carlo methods in accessing the lowest temperature scales.
Solutions of the Kondo lattice model, found via a mapping of the effective BFK
model onto the spin-boson model, have pointed to a continuous, locally critical
QPT for the 2D case \cite{Zhu:03}. By contrast, studies of the related periodic
Anderson model have predicted a strongly first-order QPT, casting doubt on the
ability of EDMFT to capture the key physics \cite{Sun:03}.

This Letter presents \textit{zero-temperature} EDMFT solutions, obtained
using a numerical renormalization-group (NRG) technique \cite{Glossop:05} to
solve the underlying BFK model. Results for both 2D and 3D spin fluctuations
point to a continuous QPT, at which the magnetic order parameter vanishes
continuously as the static susceptibility at the AFM ordering wave vector
diverges, and the terminus of PM solutions coincides with the onset
of magnetic ordering. For the topical 2D case, a concomitant divergence of the
static local susceptibility signals that Kondo physics is driven critical at
the QPT, and we are able to extract an anomalous exponent in good agreement
with experiments on CeCu$_{5.9}$Au$_{0.1}$ \cite{Schroder:00}. These results
provide important support for the notion of local quantum criticality
\cite{lcqpt,Zhu:03,Si:05}, both as a description of specific heavy-fermion
systems and, more generally, as a paradigm for novel phase transitions in
other areas of physics.

\textit{Model.}---We focus on the Kondo lattice model with Ising RKKY
anisotropy, specified by the Hamiltonian
\be
\label{latham}
\hat{H} = \sum_{i,j,\sigma} t_{ij} c_{i\sigma}^{\dagger}
    c^{\phantom{\dagger}}_{j\sigma} + \sum_{i} J \bi{S}_i \cdot \bi{s}_{c,i}
  + {\textstyle\frac{1}{2}} \sum_{i,j} I_{ij} S_i^z S_j^z.
\ee
A spin-$\frac{1}{2}$ local moment $\bi{S}_i$ at site $i$ is coupled
to the on-site conduction-electron spin $\bi{s}_{c,i}$ via an exchange
$J>0$, favoring quenching of local moments. The competing tendency towards
magnetism enters through the RKKY interaction $I_{ij}$ between
the $z$ components of the localized spins on different sites $i$ and $j$.
The tight-binding parameters $t_{ij}$ determine the conduction-band dispersion
$\epsilon_{\kk}$ and hence the density of states
$\rho(\epsilon)=\sum_{\kk}\delta(\epsilon-\epsilon_{\kk})$.

\textit{EDMFT formulation.}---We study the Kondo lattice \eqref{latham} using
EDMFT \cite{edmft}, which amounts to considering a single lattice site,
described by the impurity Hamiltonian
\begin{multline}
\label{impham}
\hat{H}_{\mathrm{BFK}} = \sum_{\kk,\sigma} \epsilon_{\kk}
     c^{\dagger}_{\kk\sigma} c^{\phantom{\dagger}}_{\kk\sigma}
   + \sum_{\qq} \w_{\qq} \phi^{\dagger}_{\qq} \phi^{\phantom{\dagger}}_{\qq}
   + J \bi{S} \cdot \bi{s}_c \\[-1ex]
+ S_z \sum_{\qq} g_{\qq} \left( \phi^{\phantom{\dagger}}_{\qq}
   + \phi^{\dagger}_{-\qq} \right) + \hloc S_z,
\end{multline}
at which a single spin $\bi{S}$ interacts with the on-site spin $\bi{s}_c$
of a conduction band and, via $S_z$, with both a dissipative bosonic bath
(representing magnetic fluctuations due to spins at all other lattice sites)
and a local magnetic field $\hloc$.
The coupling of the impurity to the nonlocal degrees of freedom is fully
specified by $J\rho(\epsilon)$ and the bosonic spectral function
$B(\w)=\pi\sum_{\qq}g_{\qq}^2\,\delta(\w-\w_{\qq})$.

Within EDMFT, the dynamical spin susceptibility is written
$\chiqw=[M(\w)+I_{\qq}]^{-1}$ in terms of a ``spin self-energy'' $M(\w)$ and
$I_{\qq}$, the spatial Fourier transform of $I_{ij}$. The key approximation in
EDMFT is neglect of the momentum dependence of both $M$ and the self-energy
$\Sigma_{\sigma}$ entering the conduction-electron Green's function
$G_{\sigma}(\kk,\epsilon)=
[\epsilon-\epsilon_{\kk}-\Sigma_{\sigma}(\epsilon)]^{-1}$. $M$ and
$\Sigma_{\sigma}$ are determined self-consistently by requiring that the
wave-vector average of the lattice correlation function equals the
corresponding \textit{local} correlation function of the impurity problem
\eqref{impham}. Thus, the local susceptibility must satisfy
\be
\cl(\w) \equiv
i \!\! \int_0^{\infty} \!\!\!\! dt \: e^{i\omega t} \bra[S_{z}(t),S_z(0)]\ket
= \int \!\! \frac{\rhoi(\epsilon) \, d\epsilon}{M(\w)+\epsilon} ,
\label{wvavg}
\ee
$\rhoi(\epsilon)=\sum_{\qq}\delta(\epsilon-I_{\qq})$ being the RKKY
density of states. The Weiss field $\czi(\w)=M(\w)-\cl^{-1}(\w)$ determines the
bosonic bath density of states through $\mathrm{Im}\czi(\w)=
\mathrm{sgn}(\w)B(|\w|)$. Analogously, $G_{0,\sigma}^{-1}(\epsilon)=
G^{-1}_{\mathrm{loc},\sigma}(\epsilon)-\Sigma_{\sigma}(\epsilon)$ determines
the band density of states $\rho(\epsilon)$. However, to study
critical properties, it is not necessary to enforce self-consistency on
$G_{\sigma}(\kk,\epsilon)$. Following \cite{Zhu:03}, we instead take
$\rho(\epsilon)=\rho_0 \, \Theta(D\!-\!|\epsilon|)$ and work with
fixed $\rho_0 J$, taking $D=1$ as the energy unit. This choice of a featureless
conduction band avoids double counting of RKKY interactions, discussed
\cite{Si:05} as the source of discrepancies between previous $T>0$ studies
\cite{Zhu:03,Si:05,Sun:03}.

For PM solutions, $\hloc\equiv 0$, and we follow solutions upon increasing the
control parameter $\delta=I/\tkn$; here, $-I\equiv I_{\bigqq}$ is the most
negative value of $I_{\qq}$, found at the AFM ordering wave vector $\qq=\bigqq$,
and $\tkn\equiv \cl^{-1}(\w=0; \delta=0)$ is the bare Kondo scale of the
pure-fermionic impurity problem. Instability of the PM phase is signaled by a
divergence of $\chiQz$ for some $\delta=\dcP$. For AFM solutions, $\hloc$ is
related self-consistently to a nonzero staggered magnetization
$\maf\equiv\bra S_z\ket$ by
\be
\hloc=-[I-\czi(\w=0)] \, \maf .
\label{hlocsc}
\ee
We follow AFM solutions from large $\delta$ down to $\delta=\dcA$, below which
the only solution is $\maf=0$.

\begin{figure}
\begin{center}
\centerline{\includegraphics[width=4.8cm,angle=270]{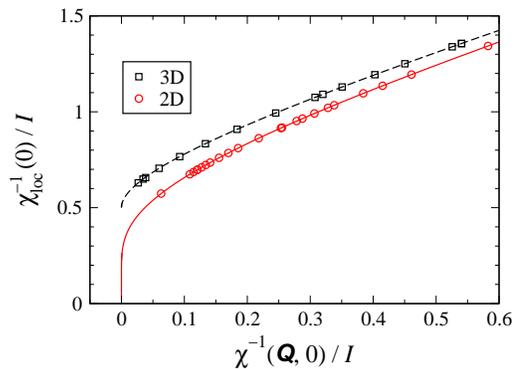}}
\caption{\label{chilocvschiQ}(Color online)
Converged values of $y\equiv[I\cl(0)]^{-1}$ vs
$x\equiv[I\chiQz]^{-1}$. In 3D (squares), Eq.\ \protect\eqref{wvavg} forces
$y=\frac{1}{2}(1+x)+\frac{1}{2}\sqrt{x(x+2)}$ (dashed line)
to approach $\frac{1}{2}$ as $x$ vanishes.
In 2D (circles), both quantities vanish together according to
$y=2/\ln(1+2/x)$ (solid line), signaling critical Kondo physics.}
\vspace*{-2ex}
\end{center}
\end{figure}

Examination of the self-consistency condition \eqref{wvavg} at $\w=0$ reveals
that the nature of the QPT depends crucially on the dimensionality of the spin
fluctuations:
In 3D, $\rhoi(\epsilon)$ has a square-root onset at its lower edge
$\epsilon=-I$. We take
$\rhoi(\epsilon)=2\sqrt{I^2-\epsilon^2}/(\pi I^2)\,\Theta(I-|\epsilon|)$, for
which Eq.\ \eqref{wvavg} gives $\cl(\w)=4\czi(\w)/I^2$, and the local static
susceptibility $\cl(0)$ remains finite when the peak susceptibility $\chiQz$
diverges (dashed line in Fig.\ \ref{chilocvschiQ}).
For 2D magnetism, $\rhoi(\epsilon)$ instead has a jump onset, caricatured as
$\rhoi(\epsilon)=(2I)^{-1}\Theta(I-|\epsilon|)$, for which Eq.\ \eqref{wvavg}
yields $\cl(\w)=(2I)^{-1}\ln\left[1+2I\chiQw\right]$. In this case, any
divergence of $\chiQz$ is necessarily accompanied by a divergence (albeit
weaker) of $\cl(0)$ (solid line in Fig.~\ref{chilocvschiQ}).

To determine where on the parametric curves shown in Fig.\ \ref{chilocvschiQ}
the EDMFT solution lies for a given $\delta$, it is necessary to solve the
full set of EDMFT equations at all frequencies. The key issue is whether, for
2D magnetism, EDMFT yields a locally critical QPT, \textit{i.e.}, whether
$\dcA=\dcP=\delta_c$ with $\cl^{-1}(\w=0;\delta=\delta_c)=0$.

\textit{Solution method.}---We solve Eq.\ \eqref{impham} using an extension
\cite{Glossop:05} of the NRG \cite{Bulla:07}. After logarithmic discretization
of the energy axis ($|\epsilon|, \, \omega =D\Lambda^{-n}$ for $n=0,1,2,\ldots$
with $\Lambda>1$), $\hat{H}_{\mathrm{BFK}}$ is recast in terms of a fermionic
and a bosonic chain, each coupled to the impurity at the first site only.
An exponential decay of tight-binding coefficients along each chain allows
iterative diagonalization of chains of increasing length. The Fock space must
be truncated, allowing a maximum of $N_b$ bosons per site, and retaining only
the lowest $N_s$ many-body eigenstates from one iteration to construct the
basis for the next.

The NRG method gives an excellent account of the universal critical properties
of the BFK model, which converge rapidly with increasing $N_b$ and $N_s$, and
are insensitive to $\Lambda$ \cite{Glossop:05}.
However, obtaining a satisfactory description of the Kondo lattice places
more stringent demands on the impurity solver.
The NRG gives the imaginary part of $\cl(\w)$ as a set of delta functions
that are broadened to recover a continuous $\cl''(\w)$ \cite{Bulla:07};
the real part $\cl'(\w)$ follows by Hilbert transformation.
When $\hloc\rightarrow 0$, $\cl(0)$ so obtained should coincide with
$-\bra S_z\ket/\hloc$, but in practice there is a mismatch between
these two values. If the self-consistently determined $\maf$, and hence
$\hloc$, vanish continuously at $\dcA$, then by rewriting Eq.\ \eqref{hlocsc}
in the equivalent form $\hloc/\maf+\cl^{-1}(0)=\chiQzi$, one sees that any
such mismatch prevents divergence of $\chiQz$, which in turn leads to
coexistence of PM and AFM solutions. It is this previously neglected mismatch
that underlies the unconventional scenarios presented in
\cite{Glossop:06,Zhu:06}, and not truncation error or a failure of EDMFT as
there suggested.

We ensure internal consistency of the NRG by applying a multiplicative
correction $1+c$ to $\cl(\w,\hloc)$, such that $\cl(0,\zp)=-\bra S_z\ket/\zp$,
where $\zp\approx 10^{-6}$ is the smallest field used in our calculations. For
$\Lambda=3$, $c\approx 0.15$ for the BFK model, very similar to the value for
the pure-fermionic Kondo model.
Within the EDMFT, the static lattice susceptibility $\chiQz$ now diverges by
construction if $\hloc\ra \zp$ self-consistently, \textit{i.e.}, if the order
parameter vanishes continuously---a scenario that we emphasize is in no way
imposed by the correction scheme.

Self-consistent EDMFT solutions are obtained as follows: For given $I$,
Eq.\ \eqref{impham} is solved using the NRG with a trial $B(\w)$ and an initial
field $\hloc>\zp$ [$\hloc=\zp$] for AFM [PM] solutions. At this point the
numerical mismatch is corrected. For AFM solutions, finding $c$ requires an
additional NRG calculation performed at $\hloc=\zp$. Then, $B(\w)$ [and for AFM
solutions, $\hloc>\zp$] is updated for use in the next loop via
Eq.~\eqref{wvavg} [and Eq.\ \eqref{hlocsc}]. Typically, 10--50 EDMFT loops
suffice to converge $B(\w)$ to within $0.0001$\% for all $\w$, though
convergence is considerably slower around the critical couplings. The solution
reached is independent of the details of the trial $B(\w)$ and, for
AFM solutions, of the initial $\hloc$.

Results are presented for $\Lambda=3$, $N_b=8$, and $N_s=300$ (requiring up
to 1 hour per EDMFT loop on a 2.2-GHz AMD Opteron CPU).
All data shown are for $\rho_0 J=0.2$, corresponding to a bare Kondo scale
$\tkn\approx 0.014$, but we find near-perfect scaling with $\tkn$ for
$\tkn\alt 0.1$. We have established in representative cases that the numerics
are converged with respect to $N_b$ and $N_s$, and that the essential physical
picture is independent of $\Lambda$.

\begin{figure}
\begin{center}
\centerline{\includegraphics[width=6cm,angle=270]{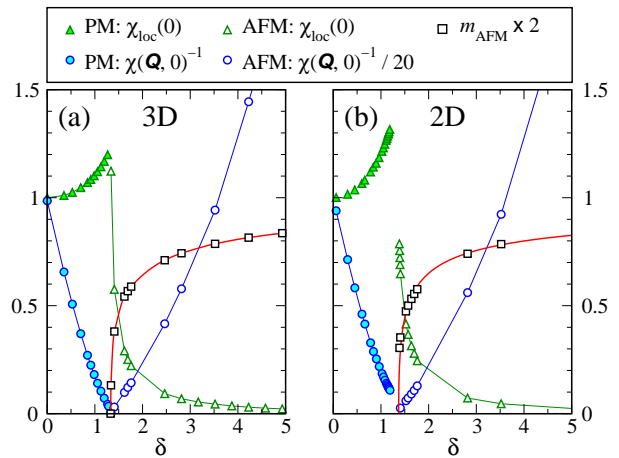}}
\caption{\label{fig:3Dand2D}(Color online)
Peak and local static susceptibilities (in units of $\tkn$) of PM and AFM
solutions, along with the order parameter $\maf$, for (a) 3D and (b) 2D
magnetic fluctuations. AFM solutions for $\delta\equiv I/\tkn\le\dcA$
(where $\maf=0$ self-consistently) coincide with PM solutions
(with $\maf\equiv 0$). Lines are guides to the eye.}
\vspace*{-2ex}
\end{center}
\end{figure}

\begin{figure}
\begin{center}
\centerline{\includegraphics[width=5.5cm,angle=270]{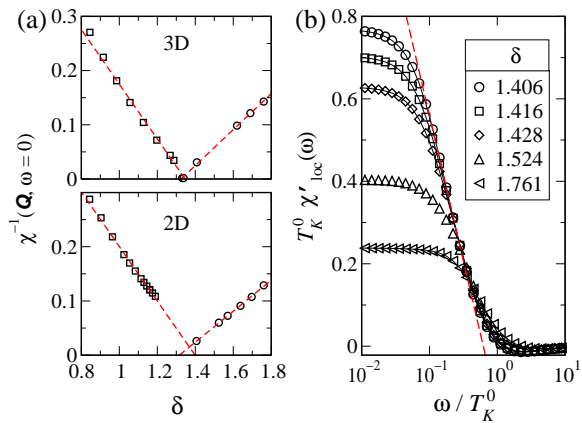}}
\caption{\label{fig3}(Color online)
(a) Extrapolation to zero of PM $\chiQzi$ (squares) and AFM $\chiQzi/20$
(circles) for 3D and 2D magnetism.
(b) Real part of $\cl(\omega)$ for five $\delta$ values approaching $\dc$
in the 2D AFM phase. The dashed line is a fit to the logarithmically singular
form that develops for $\delta\rightarrow\dc$.}
\vspace*{-2ex}
\end{center}
\end{figure}

\textit{Results for 3D magnetism.}---In this case, EDMFT has been supposed
(but not explicitly shown) to exhibit a continuous QPT of the SDW type
\cite{lcqpt}.
This picture is confirmed in Fig.~\ref{fig:3Dand2D}(a), which shows the
variation of three $T=0$ static quantities with $\delta\equiv I/\tkn$.
The inverse peak susceptibility $\chiQzi$ of the PM solutions decreases with
increasing $\delta$, and vanishes linearly at $\dcP=1.33(2)$ [see
Fig.\ \ref{fig3}(a)], beyond which no PM solutions are found. The local
susceptibility $\cl(0)$ approaches $\tkn\cl(0)=2/\dcP$ at $\dcP$, as follows
from the self-consistency (see Fig.\ \ref{chilocvschiQ}). For large $\delta$,
AFM solutions converge with nonzero $\hloc$ and $\maf$. Our data are consistent
with a continuous vanishing of $\maf$ upon decreasing $\delta$ and, provided
the internal consistency of the NRG is enforced (see \textit{Solution method}),
a concomitant divergence of $\chiQz$. Extrapolation of the lowest $\chiQzi$
values to zero yields $\dcA=1.31(2)$. For $\delta<\dcA$, AFM and PM solutions
coincide.

The large (around 30\% of $\dcA$) region of coexisting PM and AFM solutions
reported in \cite{Glossop:06} is sharply suppressed by enforcing the internal
consistency of the NRG. The reduced coexistence range of 0--5\% likely
stems from residual errors (due to discretization, truncation, Hilbert
transformation, numerical rounding, \textit{etc.}) that will tend to
destabilize the fine balance inherent to any continuous transition. We
conclude that, within numerical accuracy, the EDMFT description of the QPT
for 3D magnetism is continuous and of conventional SDW type \cite{hm}.

\textit{Results for 2D magnetism}.---Static quantities for the 2D case are
shown in Fig.~\ref{fig:3Dand2D}(b). Convergence slows markedly in the vicinity
of the transition due to the logarithmic form of the self-consistency, making
it exceedingly difficult to obtain solutions having very small $\chiQzi$.
PM-phase calculations also require a careful choice of starting
parameters to avoid flowing towards an unphysical solution \cite{Haule:02}.
Nonetheless, linear extrapolation of $\chiQzi$ to zero for each phase
[Fig.\ \ref{fig3}(a)] yields $\dcP=1.40(2)$ close to $\dcA=1.34(2)$.
The ground-state energy (not shown) provides another handle on the position
of the transition. Extrapolations of the PM and AFM energies indicate a
crossing at $\dc=1.36(1)$, consistent with $\dcA$ deduced from $\chiQzi$.

The slower convergence in 2D has prevented us from approaching the transition
as closely as in 3D, or from converging AFM solutions below $\maf\approx 0.14$
(after more than 300 EDMFT loops). However, the narrow coexistence region
(1--8\% of $\dcA$) is comparable with that in 3D, and to within our numerical
accuracy is consistent with a continuous transition. Such a QPT is necessarily
of the locally critical type \cite{lcqpt}, with $\cl(0)$ and $\chiQz$ diverging
together at $\dc$ (see Fig.\ \ref{chilocvschiQ}), where the self-consistently
determined BFK model lies at its critical point \cite{fn}. We cannot rule out
a very weakly first-order transition either in 2D or 3D, but physically this
scenario is virtually indistinguishable from the continuous case.

Support for critical local physics is provided by $\cl'(\w)$
[Fig.\ \ref{fig3}(b)], which develops a logarithmically singular form
\cite{lcqpt} $\tkn\cl'(\w)\sim(\alpha/2\dc)\ln|\w|^{-1}$ in the vicinity
of the critical coupling for $T^*_K\ll |\w| \ll \tkn$, where the effective
Kondo scale $T^*_K$ vanishes logarithmically slowly as
$\delta\rightarrow \delta_c$. We find $\alpha/2\dc=0.29(1)$ or
$\alpha=0.78(4)$, consistent with the value measured in
CeCu$_{5.9}$Au$_{0.1}$ ($\alpha=0.75$) and with that obtained in one
$T>0$ study \cite{Zhu:03}. No such anomalous exponent was found in
connection with the strong first-order behavior reported in \cite{Sun:03}.

A separate EDMFT study of the Kondo lattice has reached similar conclusions
about the nature of the QPT. In \cite{Zhu:07}, the fermionic degrees of
freedom are eliminated by mapping the BFK model onto a spin-boson model.
While this mapping introduces another level of approximation and rules out
direct comparison of the ground-state energy in the PM and AFM phases,
the spin-boson model can be solved via a purely bosonic NRG method requiring
less computational effort per EDMFT iteration, allowing investigation
closer to the critical points where convergence slows markedly. That two
complementary approaches yield essentially the same physical picture is an
important confirmation of the key results.

\textit{Summary.}---We have obtained nonperturbative, zero-temperature
solutions of the Kondo lattice model, of great present interest in connection
with heavy-fermion quantum criticality.
For magnetic fluctuations both in two and three dimensions, our results point
to a continuous transition.
In the 2D case, critical local-moment fluctuations are observed with an
anomalous exponent $\alpha \simeq 0.8$ in the dynamics that is in good
agreement with the experimentally determined value for CeCu$_{5.9}$Au$_{0.1}$.
This provides significant new evidence for local quantum criticality in
strongly correlated systems.

We thank R.\ Bulla, P.\ Coleman, S.\ Kirchner, P.\ Kumar, Q.\ Si, P.\ Sun,
P.\ W\"{o}lfle, and J.-X.\ Zhu for stimulating discussions.
We acknowledge use of the University of Florida High-Performance Computing
Center and thank C.\ Taylor for technical support.
This work was supported in part by NSF Grant DMR--0312939.


\begin{thebibliography}{25}
\bibitem{fn}
Refs.~\protect\onlinecite{Zhu:03} describe, for $\delta>\delta_c$ and
$T>0.01\tkn$, ``nominally PM'' solutions belonging to the localized phase of
the effective BFK model. We find no such solutions at $T=0$.
\end{thebibliography}
\end{document}